\documentclass[reviewcopy]{elsart}

\usepackage{adnd}
\usepackage{longtable}

\usepackage{mathptmx}

\begin{document}

\begin{frontmatter}

\journal{Atomic Data and Nuclear Data Tables}

\copyrightholder{Elsevier Science}

\runtitle{Predicted cross sections for photon-induced particle emission}
\runauthor{Rauscher and Thielemann}


\title{Predicted cross sections for photon-induced particle emission}

\author{T. Rauscher}
\ead{Thomas.Rauscher@unibas.ch}
\author{\MakeLowercase{and} F.-K. Thielemann}
\address{Departement f\"ur Physik und Astronomie, Universit\"at Basel\\
Klingelbergstr.\ 82, 4056 Basel, Switzerland}

\date{17.11.2003} 

\begin{abstract}  
Cross sections for the photon-induced particle-emission reactions
($\gamma$,n), ($\gamma$,p), and ($\gamma$,$\alpha$) are given for all
natural isotopes from Ti to Bi. The target nuclei are assumed to be
in their ground
states, except for $^{180}$Ta which is naturally occurring as the isomer
$^{180\mathrm{m}}$Ta. The cross sections are calculated in a statistical
model (Hauser-Feshbach) approach and covering an energy range from 
threshold up to 7.35 MeV above the threshold (14.7 MeV above threshold
for ($\gamma$,$\alpha$) reactions). The results are intended
to aid conception and analysis of experiments which can also be used to
test the methods involved in predicting astrophysical reaction rates for
nucleosynthesis.
\end{abstract}

\end{frontmatter}




\newpage
\tableofcontents
\listofDtables

\vskip5pc


\section{INTRODUCTION}

The importance of photodisintegration reactions for astrophysics is
manifold. In hot scenarios, such as late stellar burning phases and
the $r$-, $p$-, and $rp$-processes,
nuclei are destroyed by photodisintegration reactions
\cite{lawie,thiele02,ktw,cow,arnould,schatz}. In general,
nuclear reaction networks contain terms for both the forward and the
reverse rate, in order to account for the varying relative importance of these
reaction directions depending on temperature \cite{cow}. Recently, it has been
suggested to use Bremsstrahlung radiation \cite{mohrlett,babilon01,mohr03}
or Compton back-scattered laser
photons \cite{utso03} to measure ($\gamma$,n), ($\gamma$,p),
and ($\gamma$,$\alpha$) reactions relevant for astrophysics.
Such measurements can either be used
to test cross section predictions or to derive capture cross sections
for unstable, experimentally unaccessible targets
\cite{vogt01,belic02,vogt02,vogt03,utsoprc03}. The latter prove
important for $s$-process branchings \cite{sonn03,sonnnpa03}
which are comprised of an unstable
nucleus at which neutron capture and $\beta$-decay are competing during
the $s$-process \cite{kaepp,busso}.

In previous publications, we presented astrophysical reaction rates and
cross sections, including the ones for capture reactions
\cite{rath00,rath01}. In
principle, inverse rates and cross sections can be derived from forward
rates by detailed balance as explained in \cite{rath00}. 
In order to minimize numerical uncertainties,
reactions are usually best calculated or measured in the direction of
positive Q-value and the inverse rates are then obtained by converting
those rates. Applying this procedure to the rates and cross sections
given in \cite{rath00,rath01}, however, does not produce results which
can be directly compared to laboratory measurements as it yields
stellar cross sections $\sigma ^{*}$ where excited states in the 
target are populated
according to the given stellar temperature $T^*$
\begin{equation}
\label{csstar}
\sigma ^{*}_{\mu \nu}(E_{ij})={\sum _{\mu }(2J^{\mu }+1)\exp (-E^{\mu
}/kT^{*})
\sum _{\nu }\sigma ^{\mu \nu }(E_{ij})\over \sum _{\mu }(2J^{\mu}+1)
\exp (-E^{\mu }/kT^{*})}\quad .
\end{equation}
The initial state of the target is labeled by $\mu$, the final state of
the residual nucleus by $\nu$, the center of mass (c.m.) energy
between projectile and target by $E_{ij}$, and the Boltzmann constant by
$k$. Spin and excitation energy of a nuclear level in the target are denoted by
$J^\mu$ and $E^\mu$. The cross section $\sigma^{\mu \nu }$ for reaching
the final state $\nu$ from the initial state $\mu$ has to be known.
The conversion from the forward cross section $\sigma ^{*}_{\mu \nu}$ to
the inverse cross section $\sigma ^{*}_{\nu \mu}$ is easily made
using a simple relation (see Section 3.2 in \cite{rath00}).

Laboratory experiments
usually use targets in their ground states and thus measure
\begin{equation}
\sigma^\mathrm{lab}=\sum _{\nu }\sigma ^{0\nu }(E_{ij}) \quad . 
\end{equation}
This cannot be directly compared to the previously given stellar cross
sections $\sigma^*$. Therefore, we present new
calculations for photon-induced cross sections with targets in 
their natural excitation states.

\subsection{Method And Results}

We present ($\gamma$,n), ($\gamma$,p), and ($\gamma$,$\alpha$) cross
sections for all naturally occurring isotopes from Ti to Bi.
All targets are taken to be in their ground states, with the
exception of $^{180}$Ta which naturally occurs as the isomer
$^{180\mathrm{m}}$Ta. The cross sections are calculated with the
global statistical model NON-SMOKER
as given in \cite{rath00,rath01} (rate set FRDM). 
The reader is referred to those
references for a detailed description of the method and the used inputs.

It should be noted that the absolute energies at which the cross
sections are given are different for each target, even though the energy
grid (step width) is the same. The $\gamma$ energies $E$ are given relative to
a reference energy $E_0$ which is the particle emission threshold 
$E_0=E_\mathrm{thresh}$
unless the threshold is located at a negative
energy. In the latter case -- occurring for a number of $\alpha$ emission
cases in Table III -- the starting energy is set to zero,
$E_0=0$ MeV. The energy grids are defined slightly
differently depending on whether they are applied to neutron or charged
particle emission:
\begin{itemize}
\item
For ($\gamma$,n) cross sections in Table I, a
fine energy grid with a step width of 5 keV is used in the energy range
$E_0+0.005\leq E\leq E_0+0.245$ MeV. A larger
step width of 150 keV is used for the range $E_0+0.3\leq E\leq E_0+7.35$
MeV.
\item
For ($\gamma$,p) cross sections in Table II, the step width is 
150 keV and the considered energy range is
$E_0+0.15\leq E\leq E_0+7.35$ MeV.
\item
For ($\gamma$,$\alpha$) cross sections in Table III, the step width is
300 keV and the considered energy range is
$E_0+0.3\leq E\leq E_0+14.7$ MeV.
\end{itemize}
Thus, the targets as well as the energies cover the
relevant ranges for the $s$-process and
hot nucleosynthesis of intermediate to heavy mass
nuclei (see \cite{rath97} on how to determine the energy of the Gamow peak for
astrophysical applications).

\subsection{Limitations}

The applicability of the statistical model depends on the
excitation energy of the target obtained by photon excitation and the
nuclear level density at that energy. In most cases, the threshold
energy for particle emission is sufficiently high to guarantee a high
level density for the intermediate and heavy target nuclei considered
here. Nevertheless, the reader is advised to check the
applicability limits of the model as given in \cite{rath00}.

It should further be noted that the results are obtained with a {\em
global} approach which employs parametrizations of nuclear properties
thought to be applicable for a wide range of nuclei. This is necessary
in order to permit prediction of cross sections far off stability where
no experimental information is available. Despite of the global
character, it has been shown that the model achieves good agreement 
along the line of stability. Nevertheless, slightly higher accuracy
could have been obtained by including more experimental information and
fine-tuning to each nucleus considered here. However, this would prevent
a direct comparison to the previously published model results. For
further comments on the expected accuracy, see \cite{rath00,rath97,bao01}.

Finally, another limitation concerns the capture cross sections derived from
experiments making use of photon-induced reactions. Because they use
targets in the ground state, the resulting cross sections $\sigma^{\rm
lab}$ are only approximations to the actual stellar capture cross sections
$\sigma^*$ needed for astrophysical applications. Only one kind of
$\gamma$-transition is tested, namely the one to the ground state.
Therefore, theoretical cross sections and rates will always be
indispensable for astrophysical applications.
Nevertheless, such experiments can be a useful tool to test the stellar
particle channel as well as the ground state $\gamma$-transitions.
Since such transitions usually dominate the capture cross sections,
comparison between experimental data and the values given in this
work will provide a reasonable estimate for the accuracy of the stellar
cross sections in \cite{rath01}.

\ack
This work was supported by the Swiss NSF, grant 2000-061031.02. T. R.
acknowledges support by a PROFIL professorship (Swiss NSF
2024-067528.01).

\newpage

\section*{EXPLANATION OF TABLES}\label{sec.eot}
\addcontentsline{toc}{section}{EXPLANATION OF TABLES}

\renewcommand{\arraystretch}{1.0}

\begin{tabular*}{0.95\textwidth}{ll}
\multicolumn{2}{c}{\textbf{TABLE I. ($\gamma$,n)
Reaction Cross Sections}}\\
Target & Target nucleus $A$ for the reaction $A$($\gamma$,n)$B$.\\
&The
target is assumed to be in the ground state
except for $^{180\mathrm{m}}$Ta,\\
&where it is taken to be in the first excited state.\\
$E_0$  & Reference energy in MeV for the grid of $\gamma$-energies;\\
&in this table, it
is the energy of the neutron emission threshold.\\
$\sigma^{\rm lab}$   & Laboratory cross section
in barn for the $\gamma$-energy given in the header.\\
&The relevant energy $E$ can be computed from $E=E_0+x$,\\
&where $x$ is the energy specified in the column header.
\\
&The bracketed quantities are powers of ten,\\
&e.g.\ 2.394($-$3) means $2.394\times 10^{-3}$ barn.
\end{tabular*}
\label{tableI}

\renewcommand{\arraystretch}{1.0}

\begin{tabular*}{0.95\textwidth}{ll}
\multicolumn{2}{c}{\textbf{TABLE II. ($\gamma$,p)
Reaction Cross Sections}}\\
Target & Target nucleus $A$ for the reaction $A$($\gamma$,p)$C$.\\
&The
target is assumed to be in the ground state except for $^{180\mathrm{m}}$Ta,\\
&where it is taken to be in the first excited state.\\
$E_0$  & Reference energy in MeV for the grid of $\gamma$-energies;\\
&in this table, it
is the energy of the proton emission threshold.\\
$\sigma^{\rm lab}$   & Laboratory cross section
in barn for the $\gamma$-energy given in the header.\\
&The relevant energy $E$ can be computed from $E=E_0+x$,\\
&where $x$ is the energy specified in the column header.
\\
&The bracketed quantities are powers of ten,\\
&e.g.\ 2.394($-$3) means
$2.394\times 10^{-3}$ barn.
\end{tabular*}
\label{tableII}

\renewcommand{\arraystretch}{1.0}

\begin{tabular*}{0.95\textwidth}{ll}
\multicolumn{2}{c}{\textbf{TABLE III. ($\gamma$,$\alpha$)
Reaction Cross Sections}}\\
Target & Target nucleus $A$ for the reaction $A$($\gamma$,$\alpha$)$D$.\\
&The
target is assumed to be in the ground state except for $^{180\mathrm{m}}$Ta,\\
&where it is taken to be in the first excited state.\\
$E_0$  & Reference energy in MeV for the grid of $\gamma$-energies;\\
&this
is usually the energy of the $\alpha$ emission threshold.\\
&If the threshold energy is negative, it is written in brackets and
$E_0=0$.\\
$\sigma^{\rm lab}$  & Laboratory cross section
in barn for the $\gamma$-energy given in the header.\\
&The relevant energy $E$ can be computed from $E=E_0+x$,\\
&where $x$ is the energy specified in the column header.
\\
&The bracketed quantities are powers of ten,\\
&e.g.\ 2.394($-$3) means
$2.394\times 10^{-3}$ barn.
\end{tabular*}
\label{tableIII}

\newpage

\datatables

\setlength{\LTleft}{0pt}
\setlength{\LTright}{0pt} 


\setlength{\tabcolsep}{0.2\tabcolsep}

\renewcommand\baselinestretch{0.6}
\scriptsize 

\begin{longtable}{lcrrrrrrrrrr}
\caption{($\gamma$,n) Reaction Cross Sections}\\
\caption*{\small{See page \pageref{tableI} for Explanation of
Tables}}\\
\hline
Target&\multicolumn{1}{c}{$E_0$}&\multicolumn{2}{l}{$\sigma^{\rm lab}$ [barn] at $E_0+x$:}&&&&&&&&\\
&[MeV]&\multicolumn{1}{r}{$x=$ [keV]:}&&&&&&&&&\\
&&\multicolumn{1}{r}{5}&\multicolumn{1}{r}{10}&\multicolumn{1}{r}{15}&\multicolumn{1}{r}{20}&\multicolumn{1}{r}{25}&\multicolumn{1}{r}{30}&\multicolumn{1}{r}{35}&\multicolumn{1}{r}{40}&\multicolumn{1}{r}{45}&\multicolumn{1}{r}{50}\\
&&\multicolumn{1}{r}{55}&\multicolumn{1}{r}{60}&\multicolumn{1}{r}{65}&\multicolumn{1}{r}{70}&\multicolumn{1}{r}{75}&\multicolumn{1}{r}{80}&\multicolumn{1}{r}{85}&\multicolumn{1}{r}{90}&\multicolumn{1}{r}{95}&\multicolumn{1}{r}{100}\\
&&\multicolumn{1}{r}{105}&\multicolumn{1}{r}{110}&\multicolumn{1}{r}{115}&\multicolumn{1}{r}{120}&\multicolumn{1}{r}{125}&\multicolumn{1}{r}{130}&\multicolumn{1}{r}{135}&\multicolumn{1}{r}{140}&\multicolumn{1}{r}{145}&\multicolumn{1}{r}{150}\\
&&\multicolumn{1}{r}{155}&\multicolumn{1}{r}{160}&\multicolumn{1}{r}{165}&\multicolumn{1}{r}{170}&\multicolumn{1}{r}{175}&\multicolumn{1}{r}{180}&\multicolumn{1}{r}{185}&\multicolumn{1}{r}{190}&\multicolumn{1}{r}{195}&\multicolumn{1}{r}{200}\\
&&\multicolumn{1}{r}{205}&\multicolumn{1}{r}{210}&\multicolumn{1}{r}{215}&\multicolumn{1}{r}{220}&\multicolumn{1}{r}{225}&\multicolumn{1}{r}{230}&\multicolumn{1}{r}{235}&\multicolumn{1}{r}{240}&\multicolumn{1}{r}{245}&\\
&&&\multicolumn{1}{r}{300}&\multicolumn{1}{r}{450}&\multicolumn{1}{r}{600}&\multicolumn{1}{r}{750}&\multicolumn{1}{r}{900}&\multicolumn{1}{r}{1050}&\multicolumn{1}{r}{1200}&\multicolumn{1}{r}{1350}&\multicolumn{1}{r}{1500}\\
&&\multicolumn{1}{r}{1650}&\multicolumn{1}{r}{1800}&\multicolumn{1}{r}{1950}&\multicolumn{1}{r}{2100}&\multicolumn{1}{r}{2250}&\multicolumn{1}{r}{2400}&\multicolumn{1}{r}{2550}&\multicolumn{1}{r}{2700}&\multicolumn{1}{r}{2850}&\multicolumn{1}{r}{3000}\\
&&\multicolumn{1}{r}{3150}&\multicolumn{1}{r}{3300}&\multicolumn{1}{r}{3450}&\multicolumn{1}{r}{3600}&\multicolumn{1}{r}{3750}&\multicolumn{1}{r}{3900}&\multicolumn{1}{r}{4050}&\multicolumn{1}{r}{4200}&\multicolumn{1}{r}{4350}&\multicolumn{1}{r}{4500}\\
&&\multicolumn{1}{r}{4650}&\multicolumn{1}{r}{4800}&\multicolumn{1}{r}{4950}&\multicolumn{1}{r}{5100}&\multicolumn{1}{r}{5250}&\multicolumn{1}{r}{5400}&\multicolumn{1}{r}{5550}&\multicolumn{1}{r}{5700}&\multicolumn{1}{r}{5850}&\multicolumn{1}{r}{6000}\\
&&\multicolumn{1}{r}{6150}&\multicolumn{1}{r}{6300}&\multicolumn{1}{r}{6450}&\multicolumn{1}{r}{6600}&\multicolumn{1}{r}{6750}&\multicolumn{1}{r}{6900}&\multicolumn{1}{r}{7050}&\multicolumn{1}{r}{7200}&\multicolumn{1}{r}{7350}&\\
\hline\hline
\\[100pt]
\multicolumn{12}{c}{\small{\textit{This space intentionally left
blank}}}\\
\endfirsthead
$^{46}$Ti &13.190&2.629($-$07)&7.841($-$07)&1.511($-$06)&2.432($-$06)&3.541($-$06)&4.839($-$06)&6.326($-$06)&4.062($-$03)&5.107($-$03)&5.678($-$03)\\
&&6.073($-$03)&6.376($-$03)&6.620($-$03)&6.823($-$03)&6.998($-$03)&7.149($-$03)&7.284($-$03)&7.403($-$03)&7.511($-$03)&7.609($-$03)\\
&&7.699($-$03)&7.781($-$03)&7.857($-$03)&7.928($-$03)&7.993($-$03)&8.055($-$03)&8.112($-$03)&8.166($-$03)&8.217($-$03)&8.264($-$03)\\
&&8.310($-$03)&8.353($-$03)&8.393($-$03)&8.432($-$03)&8.469($-$03)&8.504($-$03)&8.538($-$03)&8.570($-$03)&8.602($-$03)&8.631($-$03)\\
&&8.660($-$03)&8.688($-$03)&8.714($-$03)&8.740($-$03)&8.765($-$03)&8.789($-$03)&8.812($-$03)&8.835($-$03)&8.857($-$03)&\\
&&&9.068($-$03)&1.005($-$02)&1.089($-$02)&1.162($-$02)&1.242($-$02)&1.315($-$02)&1.382($-$02)&1.446($-$02)&1.504($-$02)\\
&&1.645($-$02)&1.724($-$02)&1.835($-$02)&1.947($-$02)&2.015($-$02)&2.072($-$02)&2.122($-$02)&2.174($-$02)&2.234($-$02)&2.299($-$02)\\
&&2.401($-$02)&2.464($-$02)&2.538($-$02)&2.647($-$02)&2.709($-$02)&2.781($-$02)&2.884($-$02)&2.956($-$02)&3.006($-$02)&3.108($-$02)\\
\end{longtable}

\begin{longtable}{lcrrrrrrrrrr}
\caption{($\gamma$,p) Reaction Cross Sections}\\
\caption*{\small{See page \pageref{tableII} for Explanation of
Tables}}\\
\hline
Target&\multicolumn{1}{c}{$E_0$}&\multicolumn{2}{l}{$\sigma^{\rm lab}$ [barn] at $E_0+x$:}&&&&&&&&\\
&[MeV]&\multicolumn{1}{r}{$x=$ [keV]:}&&&&&&&&&\\
&&\multicolumn{1}{r}{150}&\multicolumn{1}{r}{300}&\multicolumn{1}{r}{450}&\multicolumn{1}{r}{600}&\multicolumn{1}{r}{750}&\multicolumn{1}{r}{900}&\multicolumn{1}{r}{1050}&\multicolumn{1}{r}{1200}&\multicolumn{1}{r}{1350}&\multicolumn{1}{r}{1500}\\
&&\multicolumn{1}{r}{1650}&\multicolumn{1}{r}{1800}&\multicolumn{1}{r}{1950}&\multicolumn{1}{r}{2100}&\multicolumn{1}{r}{2250}&\multicolumn{1}{r}{2400}&\multicolumn{1}{r}{2550}&\multicolumn{1}{r}{2700}&\multicolumn{1}{r}{2850}&\multicolumn{1}{r}{3000}\\
&&\multicolumn{1}{r}{3150}&\multicolumn{1}{r}{3300}&\multicolumn{1}{r}{3450}&\multicolumn{1}{r}{3600}&\multicolumn{1}{r}{3750}&\multicolumn{1}{r}{3900}&\multicolumn{1}{r}{4050}&\multicolumn{1}{r}{4200}&\multicolumn{1}{r}{4350}&\multicolumn{1}{r}{4500}\\
&&\multicolumn{1}{r}{4650}&\multicolumn{1}{r}{4800}&\multicolumn{1}{r}{4950}&\multicolumn{1}{r}{5100}&\multicolumn{1}{r}{5250}&\multicolumn{1}{r}{5400}&\multicolumn{1}{r}{5550}&\multicolumn{1}{r}{5700}&\multicolumn{1}{r}{5850}&\multicolumn{1}{r}{6000}\\
&&\multicolumn{1}{r}{6150}&\multicolumn{1}{r}{6300}&\multicolumn{1}{r}{6450}&\multicolumn{1}{r}{6600}&\multicolumn{1}{r}{6750}&\multicolumn{1}{r}{6900}&\multicolumn{1}{r}{7050}&\multicolumn{1}{r}{7200}&\multicolumn{1}{r}{7350}&\\
\hline\hline
\\[100pt]
\multicolumn{12}{c}{\small{\textit{This space intentionally left
blank}}}\\
\endfirsthead
$^{46}$Ti &10.345&3.997($-$19)&7.616($-$12)&9.245($-$09)&5.666($-$07)&8.770($-$06)&6.231($-$05)&2.631($-$04)&7.487($-$04)&1.584($-$03)&2.738($-$03)\\
&&4.108($-$03)&5.536($-$03)&6.883($-$03)&8.080($-$03)&9.132($-$03)&1.007($-$02)&1.092($-$02)&1.172($-$02)&1.249($-$02)&6.521($-$03)\\
&&6.574($-$03)&6.451($-$03)&6.525($-$03)&6.712($-$03)&6.890($-$03)&7.165($-$03)&7.522($-$03)&7.958($-$03)&8.475($-$03)&8.355($-$03)\\
&&8.701($-$03)&9.015($-$03)&9.249($-$03)&9.891($-$03)&1.063($-$02)&1.147($-$02)&1.229($-$02)&1.308($-$02)&1.367($-$02)&1.434($-$02)\\
&&1.517($-$02)&1.542($-$02)&1.611($-$02)&1.682($-$02)&1.701($-$02)&1.762($-$02)&1.812($-$02)&1.828($-$02)&1.869($-$02)&\\
&&&&&&&&&&&\\
$^{47}$Ti &10.462&1.342($-$20)&5.503($-$14)&2.567($-$11)&6.316($-$10)&7.658($-$09)&5.267($-$08)&2.496($-$07)&9.066($-$07)&2.694($-$06)&6.175($-$06)\\
&&1.307($-$05)&2.450($-$05)&4.343($-$05)&7.227($-$05)&1.117($-$04)&1.669($-$04)&2.398($-$04)&3.254($-$04)&4.367($-$04)&5.728($-$04)\\
\end{longtable}

\begin{longtable}{lcrrrrrrrrrr}
\caption{($\gamma$,$\alpha$) Reaction Cross Sections}\\
\caption*{\small{See page \pageref{tableIII} for Explanation of
Tables}}\\
\hline
Target&\multicolumn{1}{c}{$E_0$}&\multicolumn{2}{l}{c.s.\ [barn] at $E+x$:}&&&&&&&&\\
&[MeV]&\multicolumn{1}{r}{$x=$ [keV]:}&&&&&&&&&\\
&&\multicolumn{1}{r}{300}&\multicolumn{1}{r}{600}&\multicolumn{1}{r}{900}&\multicolumn{1}{r}{1200}&\multicolumn{1}{r}{1500}&\multicolumn{1}{r}{1800}&\multicolumn{1}{r}{2100}&\multicolumn{1}{r}{2400}&\multicolumn{1}{r}{2700}&\multicolumn{1}{r}{3000}\\
&&\multicolumn{1}{r}{3300}&\multicolumn{1}{r}{3600}&\multicolumn{1}{r}{3900}&\multicolumn{1}{r}{4200}&\multicolumn{1}{r}{4500}&\multicolumn{1}{r}{4800}&\multicolumn{1}{r}{5100}&\multicolumn{1}{r}{5400}&\multicolumn{1}{r}{5700}&\multicolumn{1}{r}{6000}\\
&&\multicolumn{1}{r}{6300}&\multicolumn{1}{r}{6600}&\multicolumn{1}{r}{6900}&\multicolumn{1}{r}{7200}&\multicolumn{1}{r}{7500}&\multicolumn{1}{r}{7800}&\multicolumn{1}{r}{8100}&\multicolumn{1}{r}{8400}&\multicolumn{1}{r}{8700}&\multicolumn{1}{r}{9000}\\
&&\multicolumn{1}{r}{9300}&\multicolumn{1}{r}{9600}&\multicolumn{1}{r}{9900}&\multicolumn{1}{r}{10200}&\multicolumn{1}{r}{10500}&\multicolumn{1}{r}{10800}&\multicolumn{1}{r}{11100}&\multicolumn{1}{r}{11400}&\multicolumn{1}{r}{11700}&\multicolumn{1}{r}{12000}\\
&&\multicolumn{1}{r}{12300}&\multicolumn{1}{r}{12600}&\multicolumn{1}{r}{12900}&\multicolumn{1}{r}{13200}&\multicolumn{1}{r}{13500}&\multicolumn{1}{r}{13800}&\multicolumn{1}{r}{14100}&\multicolumn{1}{r}{14400}&\multicolumn{1}{r}{14700}&\\
\hline\hline
\\[100pt]
\multicolumn{12}{c}{\small{\textit{This space intentionally left
blank}}}\\
\endfirsthead
$^{46}$Ti &  8.003&0.000($+$00)&3.745($-$28)&1.710($-$20)&5.647($-$16)&6.277($-$13)&1.036($-$10)&5.154($-$09)&1.140($-$07)&1.410($-$06)&1.128($-$05)\\
&&6.234($-$05)&2.268($-$04)&5.468($-$04)&9.398($-$04)&1.297($-$03)&1.628($-$03)&1.974($-$03)&1.120($-$03)&1.264($-$03)&1.415($-$03)\\
&&1.590($-$03)&1.816($-$03)&1.916($-$03)&2.059($-$03)&2.256($-$03)&2.603($-$03)&2.928($-$03)&3.250($-$03)&3.579($-$03)&3.692($-$03)\\
&&3.891($-$03)&4.110($-$03)&4.114($-$03)&4.213($-$03)&4.320($-$03)&4.204($-$03)&4.172($-$03)&4.142($-$03)&3.929($-$03)&3.801($-$03)\\
&&3.691($-$03)&3.455($-$03)&3.307($-$03)&3.189($-$03)&2.985($-$03)&2.861($-$03)&2.766($-$03)&2.607($-$03)&2.513($-$03)&\\
&&&&&&&&&&&\\
$^{47}$Ti &  8.948&0.000($+$00)&7.461($-$29)&2.183($-$21)&1.525($-$17)&6.283($-$15)&5.886($-$13)&1.035($-$11)&1.803($-$10)&2.085($-$09)&1.527($-$08)\\
&&7.996($-$08)&3.456($-$07)&1.250($-$06)&3.864($-$06)&9.970($-$06)&2.383($-$05)&5.226($-$05)&9.926($-$05)&1.798($-$04)&3.045($-$04)\\
\end{longtable}

\end{document}